# How to Make a Cylinder Roll Uphill

**Dipabali Hore[a], Abhijit Majumder[a,b], Subrata Mondal[a], Abhijit Roy[a] and Animangsu Ghatak[a]**

[a]Department of Chemical Engineering and DST unit on Soft Nanofabrication, Indian Institute of Technology, Kanpur, UP 208016, India; Tel: 0091-512-259-7146; Email: aghatak@iitk.ac.in

[b]Current address: Stem Cell Lab, Institute for Stem Cell Biology and Regenerative Medicine, Bangalore, Karnataka 560065, India.

**Abstract**

Slithering, crawling, slipping, gliding are various modes of limbless locomotion that have been mimicked for micro-manipulation of soft, slender and sessile objects. A lesser known mode is rolling which involves periodic, asymmetric and lateral muscular deformations. Here we enable an elastomeric cylinder of poly(dimethylsiloxane) to roll on a substrate by releasing small quantity of a solvent like chloroform, toluene, hexane, heptane and so on, which swells differentially a portion of the cylinder, but evaporates from portion of it which remains exposed to the atmosphere. In a dynamic situation, this asymmetric swelling-shrinking cycle generates a torque which drives the cylinder to roll. The driving torque is strong enough that the cylinder can roll up an inclined plane, within a range of inclination, its velocity even increases. The cylinder can even drag a dead weight significantly larger, ~8-10 times its own weight. A scaling law is derived for optimizing the rolling velocity.

**Introduction:**

While practical examples of chemo-mechanical devices are rather few and limited [1], in nature, instances of directed chemo-mechanical motion are abundant and can be found for all different length scales. Incorporation of these mechanisms into microfluidics, micro and nano-electro mechanical systems (MEMS & NEMS), drug delivery and for directional motion and transport are expected to enhance the design efficiency of these systems. Creating directed motion in micro-systems is not, however, easy, as it is constrained by the availability of micro-fabrication techniques; importantly, transformation of energy into motion becomes increasingly difficult with decrease in the dimension of the systems. For example, electro-magnetic [2-5] or pneumatic signals [6], which are often used to impart motion in micro-systems, are limited by their inherent needs of wires, tethers or on-board energy supply arrangements. Therefore continuous efforts have been made to design systems which can convert chemical energy directly into mechanical energy without any intermediate step or too many motion-bearing parts. The efforts to impart chemo-mechanical motion in microsystems include photo-, electro-, thermo- or pH regulated motion of soft gels and polymers [9-14]. For example, asymmetric illumination of light induces photochemical changes in the surface chemistry of porous polymeric microstructures which give rise to a substantial asymmetric volume change resulting in motion of the structure [15]. Phase transition is also achieved by application of electric field across a polyelectrolyte gel [16]. Asymmetric vibration [17] too generates motion in soft bodies. Devices based on biological systems show that directed motion can be induced by using, for example, adenosine triphosphate (ATP) powered biomolecules such as motor protein. Microtubules undergo directed motion when kinesin molecules, powered through hydrolysis of ATP molecules, walk on the microtubules with discrete 8 nm steps [18-19]. Functionalization of microtubules with magnetic nanoparticles followed by application of external magnetic fields also gives rise to directed motion [20]. In the present work, we demonstrate a novel strategy to impart rolling motion into elastomeric micro-cylinders using partial swelling-shrinking cycle. Here, we apply definite volume of organic solvents at one side of the cylinder that remains adhered to a surface. The solvent causes more swelling of that side compare to the other side of the cylinder. This differential swelling coupled with the adhesion with the rigid substrate breaks the stress-symmetry and results in a torque which imparts a rolling motion to the cylinder. We have studied the effect of geometric parameters like cylinder diameter and its length and the material properties of the solvent and that of the cross-linked network on the linear velocity of the cylinder and the distance travelled by it. We have shown that when the cylinder is placed on an inclined plane, the solvent tends to accumulates more on its rear side, as a result of which, the cylinder climbs up the plane. Within a range of angle of inclination, the rolling velocity of the cylinder was found to be even higher than that on a horizontal plane. In fact the cylinder could even carry some load while rolling up the inclined plane.

**Materials and Method:**

**Materials:** Poly(dimethylsiloxane) (sylgard 184,Dow corning product) (PDMS) was used for the preparation of elastomeric cylinders; organic liquids: chloroform, hexane, toluene, heptanes, triethylamine (TEA) and cyclohexane were used as solvents in the experiment. Surgical needles of various bore diameters were used as the template for generating these cylinders. Microscope glass slides were used as substrates for carrying out rolling of cylinders. The substrate was coated with self assembled monolayer (SAM) of Octadecyltrichlorosilane(OTS) molecules to reduce adhesion with the cylinder.

**Generation of PDMS micro-cylinders:** PDMS precursor liquid mixed with the crosslinking agent (5:1 to 20:1 by weight) was sucked into a cylindrical mold, e.g. steel surgical needles using surgical syringes following which, it was cross-linked at $80^0 C$ for $1$ hour. The crosslinked PDMS cylinders were then gently pulled out from the needles. Cylinders of diameters: $d = 330 - 870 \mu m$ and length: $l = 10$ mm were generated by this method.

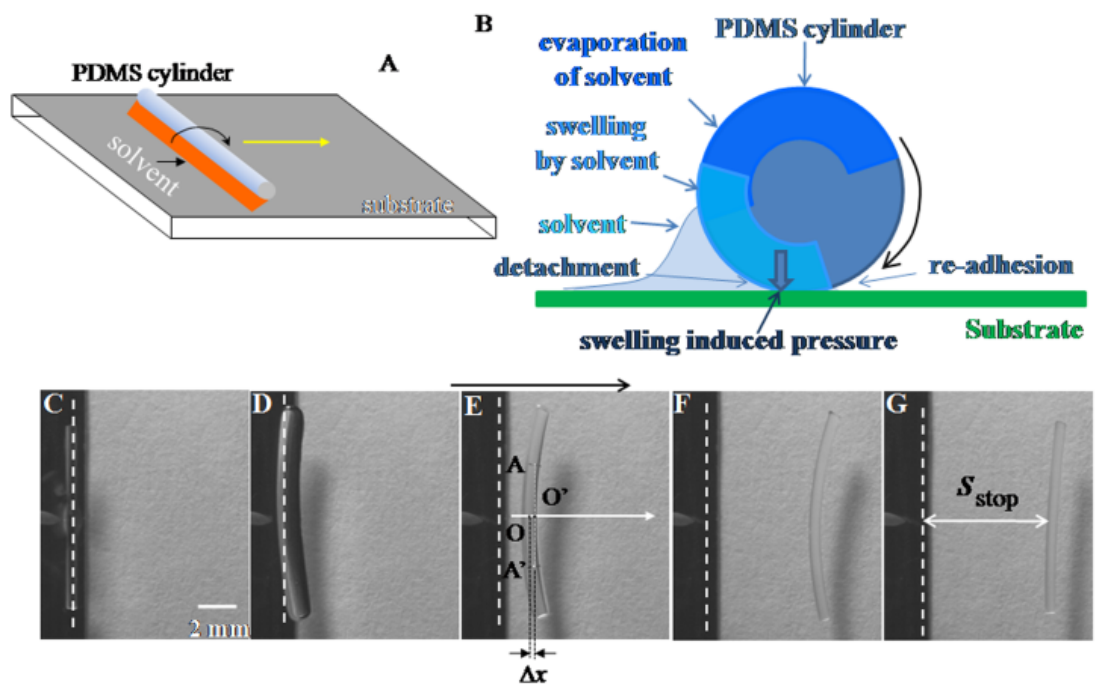

**Figure 1: A.** Schematic of the experiment in which a PDMS cylinder is placed on a substrate. Small quantity of a solvent is released from a motorized dispenser to a side of the cylinder which triggers the cylinder to roll. **B.** The magnified side view depicts the mechanism of rolling of the cylinder. **C-G.** Video micrographs depict a typical sequence of rolling of a cylinder of diameter $d = 550$ μm and length $l = 10$ mm following release of $4$ μl of chloroform. Image C: initial state of the cylinder at the point of release of solvent; D to G: its position at intermediate times: 8, 13, 19, 25.3 sec respectively.

| Solvent | Boiling Point (°C) | Vapor Pressure at 27°C (kPa) | Surface Tension (mN/m) | Solubility Parameter (cal$^{1/2}$cm$^{-3/2}$) | Swelling Ratio | Density (gm/cm$^3$) |
|---|---|---|---|---|---|---|
| Acetone | 56.53 | 33.33 | 25.20 | 9.9 | 1.06 | 0.785 |
| Chloroform | 60 | 26.14 | 27.5 | 9.2 | 1.39 | 1.489 |
| *n*-Hexane | 69 | 20.25 | 18.4 | 7.3 | 1.35 | 0.655 |
| Ethanol | 78.4 | 8.75 | 22.39 | 12.7 | 1.04 | 0.789 |
| Cyclohexane | 80.7 | 13.17 | 24.9 | 8.2 | 1.33 | 0.779 |
| TEA | 90 | 9.13 | 23.4 | 7.5 | 1.58 | 0.730 |
| *n*-Heptane | 98 | 6.07 | 20.1 | 7.4 | 1.34 | 0.680 |
| Water | 100 | 3.51 | 72.86 | 23.4 | 1.00 | 1.000 |
| Toluene | 111 | 3.8 | 28.4 | 8.9 | 1.31 | 0.862 |

**Table 1: Table shows different solvents used in the experiment and their physical properties. Here vapor pressure is calculated using Antoinne Equation. Solubility parameter and swelling ratio data are taken from reference 21 of the manuscript. The surface tension data of liquids were obtained from reference 22. The density data were taken from reference 25.**

**Experiment:** Figure 1 (A and B) depicts the schematic of the experiment in which a slender poly(dimethylsiloxane) (PDMS) cylinder is placed on a microscope glass slide coated with self assembled monolayer of octadecyltrichlorosilane (OTS) molecules. The surface of these cylinders has roughness of that of the inner surface of the needles; from Atomic Force Microscopy (supporting figure S1) measurement the root mean square roughness is estimated to be ~200nm. When such a cylinder is placed on a glass slide, it does not form a continuous contact as happens with a smooth cylinder, instead, remains weakly adhered to it. We use solvents like chloroform, n-hexane, n-heptane, toluene and tri-ethylamine, with various properties like boiling point ($T$), vapor pressure ($p$), surface tension ($\gamma$), density ($\rho$) and solubility parameter ($\delta$) as presented in Table-1. The last property is a measure of the intermolecular interactions and defines swellability of the solid, e.g. PDMS swells most in solvents with $\delta$ similar to that of PDMS: $\delta_{PDMS} = 7.3$ [21]. A small quantity of solvent is released at one side of the cylinder using a micro-pipette following which the rolling motion of the cylinder sets in. The substrate and the ambient temperature are maintained at $26.5 \pm 0.5^0 C$ in all experiments.

**Results and Discussion:**

Optical micrographs in figure 1C-G depict the sequence of rolling of a cylinder of diameter $d = 550 \mu m$ and length $l = 10$ mm on a horizontal glass slide. Here, 4 µl of chloroform is released at the rear side of the cylinder, close to its centre along the axial length. If the solvent is added near to one of its edges, the initial movement of the cylinder becomes erratic and unstable; in some cases the cylinder exhibits out of plane bending. However, the liquid soon gets evenly distributed along the length, following which uniform rolling motion of the cylinder sets in.

The schematic of the side view of a typical cylinder in its rolling state is depicted in figure 1B which captures the mechanism of rolling. Here, the quantity of solvent is insufficient to entirely submerge the cylinder, hence it does not swell uniformly everywhere but swells more where it is in contact with the solvent compared to the portion away from it. The solvent evaporates from the portion of the cylinder exposed to the atmosphere resulting in shrinkage. This asymmetric swelling results in bending of the cylinder (figure 1D-F and supporting figure S2), because of which, its center of mass moves ahead of its position in the straight condition as shown in figure 1E. The new position O' of center of mass remains slightly ahead of the old position O and the rest of the portion of the cylinder tends to catch up with it. The localized swelling of the cylinder at the solvent side exerts also the swelling pressure on the substrate; the reaction force, in turn generates a torque which then propels the cylinder to rotate forward. In a dynamic situation, the cylinder always drags with it a pool of the solvent so that it remains locally swelled at the rear side and solvent evaporates from the free surface. As a result, the driving torque remains maintained for rotation along a particular direction. The cylinder continues to roll till it straightens out (figure 1G) following evaporation of all the solvent. A cylinder with smooth outer surface is however not set to roll in similar experiment as it adheres strongly with the substrate.

Figure 2A depict the distance $s$ travelled by a cylinder of diameter $d = 550$ μm for various quantity of chloroform. Here $t = 0$ defines the time at which the solvent is first released, following which the linear translational velocity of the cylinder increases over a time $t_{lag}$ from zero to a final linear velocity $v$. During this transition, the cylinder rolls but with complete to partial slippage on the substrate. The cylinder rolls with velocity $v$ for the rest of its travel time, $t_{roll}$. The time $t_{lag}$ decreases with the decrease in the initial quantity of the solvent, for very small volume of the solvent, almost no lag period is found. In figure 2B we show that rolling velocity $v$ varies non-monotonically with the quantity $Q$ of solvent. For example, in experiments with chloroform, small quantity of solvent: $Q = 0.5 - 2$ μl evaporates even before it spreads along the length of the cylinder. The resultant torque is small enough to engender only a small velocity $v \approx 0.5$ mm/sec. Velocity $v$ increases with $Q$ finally attaining a maximum velocity at an intermediate $Q = 4$ μl. For large $Q = 15 - 43$ μl, the solvent deluges the cylinder nearly completely, resulting in reduced asymmetry in swelling and smaller velocity of the cylinder.

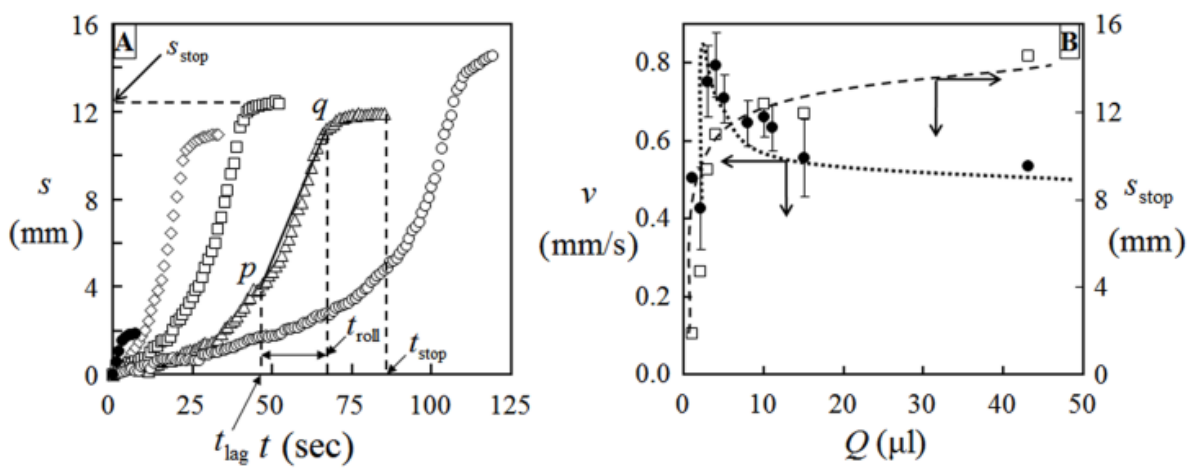

**Figure 2.** A. Plot shows distance rolled by the cylinder with time for different volume of chloroform. Symbols ●, ◇, □, △ and ○ represent $Q = 1, 3, 4, 15$ and $43$ μl respectively. $t_{lag}$ and $t_{roll}$ represent the lag and constant velocity period respectively. Slope of $s$ vs. $t$ data within time $t_{roll}$ i.e. that of line $pq$ yields the linear velocity $v$ of the rod. $t_{stop}$ and $s_{stop}$ represent respectively the total time and net distance of travel by the cylinder. B. Plot shows velocity $v$ and distance $s_{stop}$ rolled by the cylinder against volume $Q$ of the solvent. The lines are guide to the eye.

In figure 3A we plot the linear velocity $v$ of cylinders against diameter $d$ for a fixed volume of $4$ μl of solvent released at the vicinity of the cylinders. For all solvents, $v$ increases nearly linearly with diameter till a maximum diameter is reached beyond which enough torque is not generated to drive the cylinder; it then deforms in bending mode. For chloroform this threshold diameter is found to be $d_{max} \sim 1260$ μm. The data reveal also the combined effect of different solvent properties. For example, chloroform (■) and toluene (◇) have similar values of solubility parameter: $\delta_{toluene} = 8.9$, $\delta_{chloroform} = 9.2$ and surface tension: $\gamma_{toluene} = 28.4$ mN/m and $\gamma_{chloroform} = 27.5$ mN/m but different boiling point or vapor pressure: $T_{chloroform} = 60\,^{\circ}\mathrm{C}$, $T_{toluene} = 111\,^{\circ}\mathrm{C}$. As a result, at $27\,^{\circ}\mathrm{C}$ the liquid with lower boiling point i.e. chlorofrom has higher vapor pressure than toluene, enabling it to evaporate faster. A higher rate of evaporation leads to faster deswelling and higher rotational velocity of the rod. Hence, chloroform generates a larger rolling velocity of the cylinder than toluene. Similarly, n-hexane (◆) and chloroform (■) have similar boiling temperature, $T_{hexane} = 69\,^{o}C$ and $T_{chloroform} = 60\,^{o}C$, but different surface tension: $\gamma_{chloroform} = 27.5\,mN/m$, $\gamma_{hexane} = 18.4\,mN/m$ and solubility parameter: $\delta_{chloroform} = 9.2$ and $\delta_{hexane} = 7.3$. Because of lower surface tension and higher solubility, hexane spreads faster onto the surface of the rod, so that large surface area gets exposed to atmosphere which in turn increases the evaporation rate. Consequently, the swelling-shrinking cycle goes on at faster rate enabling the cylinder to roll at a larger velocity. These observations suggests that solvents with similar $\gamma$, $p$ and $\delta$ should impart similar rotational speed to the cylinders, which is corroborated by experiment with n-heptane (□) and triethylamine (○). These solvents have similar values of solubility parameter: $\delta_{heptane} = 7.4$ and $\delta_{TEA} = 7.5$, boiling point: $T_{heptane} = 98\,^{o}C$ and $T_{TEA} = 90\,^{o}C$ and surface tension: $\gamma_{heptane} = 20.1$ mN/m and $\gamma_{TEA} = 23.4$ mN/m and they impart very similar velocity to the cylinders as shown by figure 3A. We examine also the effect of cylinder elastic modulus which can be varied by varying the crosslinking density of the elastomer. We have used different weight fraction of the crosslinker: 5% - 20% by weight in order to generate cylinders with shear modulus $\mu$ varying from 0.5-1.6 MPa. The shear modulus was quantified using Johnson-Kendall-Robert (JKR) type contact mechanics experiments [26] on these networks. The inset in figure 4A shows that with hexane, velocity $v$ increases linearly with $\mu$. The rolling velocity of ~2 mm/sec thus achieved, exceeds most previous examples by at least an order of magnitude [1,10].

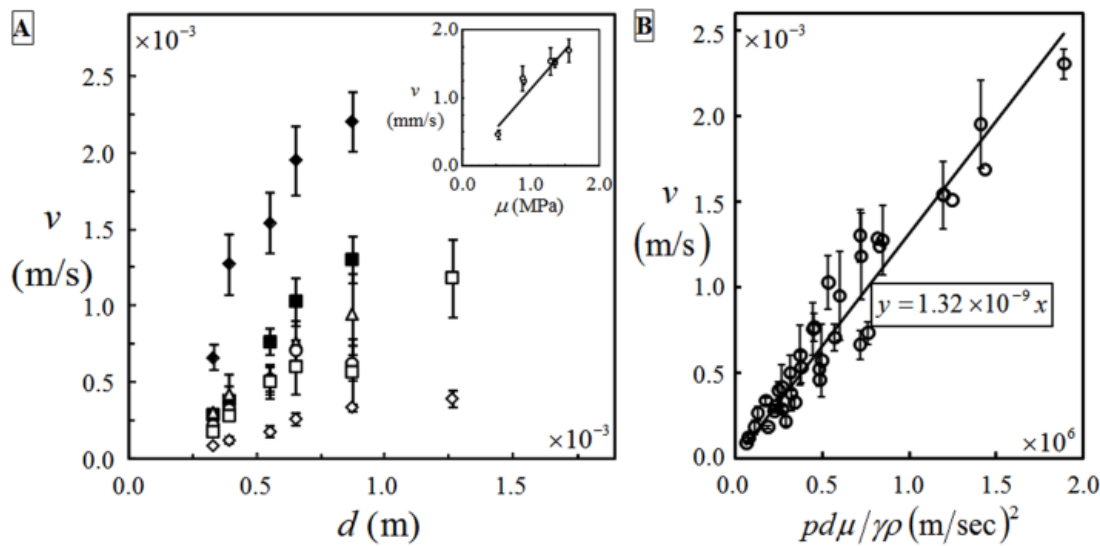

**Figure 3.** A. The plot of linear velocity $v$ of PDMS micro-cylinders of length $l = 10$ mm and modulus $\mu = 1.3$ MPa against their diameter for 4 $\mu l$ of various solvents. The symbols ◆, ■, ∆, ○, ◇ and □ represent solvents: *n*-hexane, chloroform, TEA, cyclohexane, toluene and *n*-heptane respectively. The inset shows the velocity of cylinders of diameter $d = 550$ μm and length $l = 10$ mm plotted against their shear modulus. The cylinders are driven by 4 $\mu l$ of hexane. The error bars signify the standard deviation of data obtained from several experiments using different PDMS cylinders. B. Plot of linear velocity $v$ of cylinders of different cross-linking density and diameter for solvents with different vapor pressure and surface tension with respect to the quantity $pd\mu/\gamma\rho$. The symbol ○ representing data from all solvents can be fitted to a master straight line represented by the solid line.

The above observations can be rationalized by considering that during one rotation of the cylinder, strain energy $\sim \mu\varepsilon^2 \frac{\pi d^2 l}{4}$ is released because of its swelling and shrinking, which is used in wetting a surface area $vtl$ by the liquid of surface tension $\gamma_l$. Here $\varepsilon$ is the characteristic strain, which can naively be associated to the swelling of the cylinder as $\varepsilon = (d'/d - 1)$, where $d'$ is its swelled diameter.

Balance of these two energies yield an expression for the velocity of rolling, $v \sim \frac{\pi d^2 \mu}{4\tau\gamma_l}(d'/d - 1)^2$ in terms of shear modulus $\mu$ and diameter $d$ of the cylinder, surface tension $\gamma_l$ of the solvent and a characteristic time scale $\tau$ during which the cylinder rolls over one complete rotation. The timescale $\tau$ is a measure of the rate of evaporation of the solvent from the curved surface of the cylinder which increases with vapor pressure $p$ of the solvent but decreases with the diameter of the cylinder. $\tau$ can be obtained from simple mass balance, e.g. by equating the rate of loss of mass from the cylinder to the rate of evaporation of the solvent to the surrounding. For example, through time $\tau$, the swollen cylinder of diameter $d'$ decreases to $d$, so that, the average rate of solvent evaporation can be written as $\sim \frac{\rho \pi d^2 l ((d'/d)^2 - 1)}{4\tau}$ where $\rho$ is the density of the solvent. The solvent evaporates because of difference in its vapor pressure at the surface of the cylinder and its partial pressure at the surroundings. As the cylinder rolls into a surrounding which is fresh and devoid of any solvent, we assume that the partial pressure of the solvent in the surroundings is negligibly small. Then the rate of evaporation of the solvent is proportional to the vapor pressure and the curved surface area of the cylinder and can be written as $\kappa \pi d l p$; here the proportionality constant $\kappa$ has a unit of (m/sec)$^{-1}$ and is a measure of the rate of evaporation of solvent per unit difference in pressure, per unit surface area of the cylinder. Equating these two quantities result in an expression for the characteristic time, $\tau \sim \frac{\rho \pi d ((d'/d)^2 - 1)}{4 p \pi \kappa}$. Substituting this expression for $\tau$ into that of $v$ and noting that for small extent of swelling, $((d'/d)^2 - 1) \sim (d'/d - 1)^2$, we obtain the following simple scaling relation for the translational velocity of the cylinder: $v \sim \kappa \frac{d p \mu}{\gamma \rho}$. This expression shows that the velocity of rolling of the cylinder should increase with the diameter of the cylinder and its shear modulus as observed in experiments. However, assumption of small swelling excludes the possibility of non-linear dependence of $v$ on $d$ which is observed at large swelling of small diameter cylinders. In Figure 3B, the data of linear velocity $v$ from experiments using cylinders of different diameters and modulus and different solvents indeed all fall on a single master lime, with the result, $v = 1.32 \times 10^{-9} \frac{pd\mu}{\gamma\rho}$ m/sec.

We will now show that by the above mechanism the cylinder can be enabled to work against an external load. For example, the cylinders can be made to roll up a stiff inclined plane as shown by a typical sequence of video micro-graphs A-C in figure 4 (see also movie S1). Here, chloroform is the driving solvent. In figure 4D we plot $v$ against $\theta$ for three different solvents: chloroform, TEA and n-heptane. For each case, $v$ increases till a threshold $\theta$ is reached, beyond which it decreases. Two opposing forces act on the cylinder: gravitational pull tends to drag the liquid towards the rear side of the cylinder and even down the inclined plate so that a smaller quantity of solvent is available for generating the asymmetric swelling of the cylinder. Therefore, the effect of differential swelling of the cylinder gets more pronounced as compared to what is attained on a horizontal plane. This effect is similar to experiments on a horizontal plane in which a smaller quantity of solvent generates larger rolling velocity (Figure 2B).

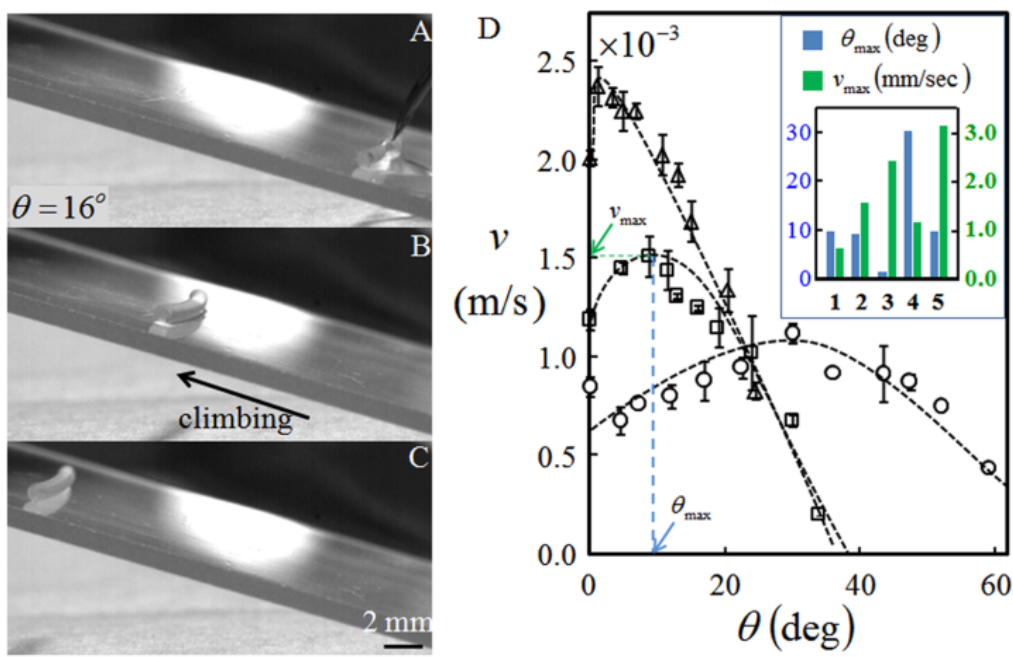

**Figure 4: Video micrographs (A)-(B) show a chloroform driven PDMS cylinder of diameter $d = 870\,\mu\text{m}$ rolling up a plane, inclined at an angle $16^0$ to the horizon. The images (A-C) are obtained at time $t = 0$, $6.3$ and $12$ sec respectively. D. The velocity $v$ of a cylinder of $d = 870\,\mu\text{m}$ is plotted against angle of inclination $\theta$ of a substrate. Symbols $\triangle$, $\square$ and $\bigcirc$ represents solvents: chloroform, TEA and n-heptane respectively. Solid lines are guide to the eye. The bar chart (1-5) at the inset shows the $\theta_{max}$ and $v_{max}$ data for solvents toluene, TEA, chloroform, heptanes and hexane respectively.**

However, beyond $\theta_{max}$ the gravitational pull on the solvent and the cylinder dominates so that the cylinder slows down, eventually it even slides down the inclined plane. The data with different solvents (figure 4D) show that velocity decreases more for solvents with larger density. For chloroform ($\rho = 1.489$), the rolling velocity is larger than for TEA ($\rho = 0.73$) and n-heptane ($\rho = 0.68$) on a horizontal surface, yet it decreases to zero within $0^0 < \theta < 30^o$. However, for n-heptane the cylinder continues to roll for $\theta$ as high as $55^o$ suggesting that a solvent with smaller density is possibly more appropriate for driving a cylinder uphill. The bar chart at the inset of figure 4D shows the angle of inclination, $\theta_{max}$ at which the rolling velocity maximizes for different solvents and the maximum velocity, $v_{max}$ attained. The data show that $\theta_{max}$ can be as high as $30^o$ for heptane.

While on an inclined plane the cylinder ascends against its own gravitational pull, the sequence of images Figure 5A-C (See also movie S2) shows that on a horizontal surface it can comfortably drag a dead weight much higher than its own. Here, the rolling experiment is carried out using a PDMS cylinder with an elongated object like a ball-pin placed on it. The cylinder is found to pull the object forward, on application of a solvent at its rear side. In other experiments in which a smaller object is needed to be transported, it is placed on the front side of the cylinder which now pushes it forward, signifying that the cylinder can both pull and a push an object. Powered by 2μl of n-hexane, a cylinder of weight 4mg drags an 18mg object. The data in figure 5D shows that with increase in weight of the object, the rolling velocity decreases linearly, but only slightly, till a critical weight is reached, beyond which it decreases catastrophically to zero. In fact the cylinder can comfortably drag an object 8 to 10 times its own body weight.

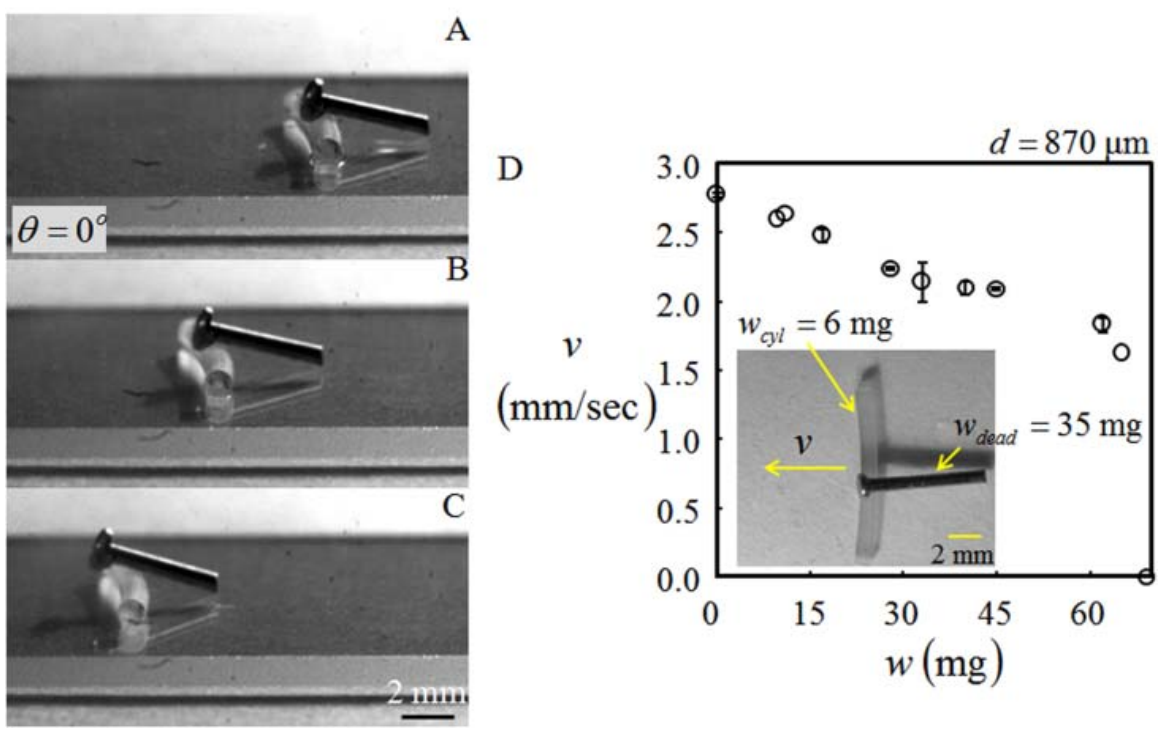

**Figure 5. A-C. The sequence of images captured at $t = 0$, $1.43$ and $2.87$ sec respectively show a cylinder of weight $4$ mg dragging a dead weight of $18$ mg on a horizontal surface. Figure D shows typical data of rolling velocity for a PDMS cylinder of diameter, $d = 870$ μm and weight, $w_{cyl} = 6$ mg transporting objects of different weights on application of $Q = 2$ $\mu l$ of n-hexane. With increase in weight of the object, the rolling velocity decreases linearly, but only slightly, till a critical weight is reached, beyond which it decreases catastrophically to zero. Thus, the cylinder comfortably carries an weight 8 to 10 times its own body weight.**

**Summary:**

To summarize, here we describe a novel autonomous rolling motion of a soft elastomeric cylinder which is found to occur via a fairly complicated yet coordinated sequence of events: asymmetric swelling of the cylinder with release of small quantity of solvent, buckling of the cylinder with center of mass moving forward and debonding of the cylinder from the substrate at its rear and adhesion onto it at the front and all happening continuously in a dynamic co-ordinated cycle till all the solvent gets evaporated. Unlike previously explored mechanisms for inducing locomotion in limbless soft bodies, here we see several counter-intuitive observations: increase in rolling velocity with decrease in quantity of solvent released and increase in inclination of the substrate. We show also that the cylinder is capable of carrying a cargo much larger than its own bodyweight which opens up the possibility of many practical applications, for example, in design of soft robotic components, soft actuators and transport and delivery of small objects, specially, in the area of MEMS and NEMS and so on.

**Acknowledgement:**

AG acknowledges IRHPA grant from Department of Science and Technology, Government of India for this work.